# Simple approach to model the strength of solid-solution high entropy alloys in Co-Cr-Fe-Mn-Ni system

Ali Shafiei

Metallurgy Group, Niroo Research Institute (NRI), Tehran 14665-517, Iran
E-mail: alshafiei@nri.ac.ir, Tel: +98 (21) 88074187

**Abstract**

A simple fitting approach is introduced for modeling the strength (hardness) of quaternary and quinary face-centered cubic (fcc) solid solution high entropy alloys (HEAs) in Co-Cr-Fe-Mn-Ni system. It is proposed that the strength of solid solution HEAs could be modeled by a polynomial equation where experimental data are used for finding the coefficients of polynomial. It is observed that the proposed polynomial could model the strength of solid solution HEAs very well. Effects of constituent elements on the hardness of quinary Co-Cr-Fe-Mn-Ni alloys are investigated; the results indicate that the strength of alloys decreases with increasing the Fe content. The softening effect of Fe is explained by considering its effect on decreasing the shear modulus of alloys. Furthermore, the effects of parameters enthalpy of mixing and valence electron concentration on the strength of HEAs are investigated. The results show that the enthalpy of mixing has a noticeable effect on the hardness of quinary Co-Cr-Fe-Mn-Ni alloys and the strength increases with decreasing the enthalpy of mixing. Furthermore, the results show that hardness of quinary Co-Cr-Fe-Mn-Ni alloys increases with increasing the parameter valence electron concentration.

**Keywords**: hardness; microstructure; metal and alloys; modeling; high entropy alloys

# 1. Introduction

Multi-principal element alloys or multicomponent alloys with at least three principal elements (e. g. CoCrNi or CoCrFeMnNi) have received wide attentions recently because they may have enhanced properties in comparison with traditional alloys which are usually based on one dominant element [1-2]. While these alloys contain several principal elements, sometimes they crystallize in single-phase solid solutions with simple crystal structures. The occurrence of alloys with single-phase solid solutions is attributed to high configurational entropies of random solid solutions, and that is why these alloys are usually referred as medium entropy (alloys with three or four principal elements) or high entropy (alloys with at least five principal elements) alloys [3-4].

Because high entropy alloys (HEAs) contain several principal elements, the composition domain from which an alloy can be selected is very large. Therefore, the exploration of the whole composition domain to identify alloys with desired properties is very challenging. To overcome this problem, many research works have been focusing on developing models for predicting the microstructure or properties of HEAs [5-9]. Such models can facilitate the exploration of the whole composition domain leading to the identifying of HEAs with desired properties. Different strategies have been used for predicting the microstructure or properties of HEAs [5-9]. Most of the approaches are based on thermodynamic simulations or *ab initio* calculations. Although these approaches were successful in predicting the structure or mechanical properties of HEAs, they have their own limitations.

The most comprehensive model which is usually used for estimating the solid solution hardening in HEAs, is the model derived by Varvenne et al. [9]. In the developed model by Varvenne et al. each elemental alloy component was assumed as a solute which is embedded in an effective solvent [9]. The solute was then considered as a local concentration fluctuation which can interact with dislocations leading to the hardening of the alloy [9]. Although the model has been successfully used in several research works for predicting the strength of solid-solution high entropy alloys, the theory has some significant drawbacks as stressed in [10]. Furthermore, it is reported very recently that the Varvenne's model (used in a reduced form) fails to predict the changes of strength in solid solution high entropy alloys $(AuNiPdPt)_{1-x}Cu_x$ [10]. Therefore, it may be concluded that a model which can accurately predict the strength of high entropy alloys is still lacking.

The objective of the present work is to introduce a new approach for modelling the mechanical properties of solid solution HEAs. Although the introduced approach (data-based modeling and fitting) does not explain the parameters and mechanisms involved in strengthening of HEAs, it can be considered as a straightforward method for assessing the strength of solid solution HEAs. Therefore, the proposed approach can be considered as a valuable tool for designing HEAs. Quaternary and quinary fcc solid solution HEAs in the Co-Cr-Fe-Mn-Ni system (Cantor alloy based system) are investigated in the present work, however it is believed that the approach can be used for other solid solution systems as well.

## 2. Methodology

### 2. 1. Experimental data

The experimental dataset which is shown in Table 1 is used in the present work. The data in Table 1 are gathered from a recently published paper by Bracq et al [11]. It was reported that all of the alloys listed in Table 1 have a single phase fcc solid solution after annealing [11] except alloy #1 which contained small amounts (4% volume fraction ) of a bcc phase. Because nanoindentation tests were performed on the center of single grains, the role of grain-boundary strengthening on hardness values can be ignored. So, the hardness values can be directly attributed to the frictional stress (or the intrinsic lattice resistance to dislocation motion) and the strengthening contribution from solid solution hardening [9-12]. It should be noted that the frictional stress and solid solution hardening both can be considered as a function of chemical composition [9-12]. Furthermore, it should be noted that because same procedures (casting and heat treatment) were used for the preparation of alloys in Table 1, therefore, the role of the microstructure on the hardness values can be ignored and the hardness values can be directly attributed to chemical compositions. The composition domain which is covered by alloys in Table 1 is schematically shown in Figure 1. It can be seen that the dataset in Table 1 covers a noticeable part of the composition domain.

Table 1. Experimental data used in the present work [11]

| Alloy | Chemical composition (at. %) | | | | | hardness H (GPa) | lattice parameter $a$ (Å) |
|---|---|---|---|---|---|---|---|
| | Co | Cr | Fe | Mn | Ni | | |
| 1 | 0 | 23.8 | 24.9 | 25.8 | 25.5 | 2.12 | 3.622 |
| 2 | 10 | 23.1 | 22.5 | 22.5 | 21.9 | 2.5 | 3.609 |
| 3 | 20 | 20.3 | 20 | 20.3 | 19.4 | 2.52 | 3.601 |
| 4 | 29.7 | 18.1 | 17.5 | 17.7 | 17 | 2.53 | 3.593 |
| 5 | 49.6 | 12.8 | 12.5 | 12.7 | 12.4 | 2.74 | 3.576 |
| 6 | 25.1 | 0 | 25 | 25.2 | 24.7 | 2.07 | 3.603 |
| 7 | 23.7 | 5.2 | 23.9 | 24.2 | 23 | 2.41 | 3.602 |
| 8 | 21.2 | 15.3 | 21.3 | 21.6 | 20.6 | 2.53 | 3.601 |
| 9 | 18.7 | 25.5 | 18.7 | 18.9 | 18.2 | 2.5 | 3.605 |
| 10 | 24.9 | 25.6 | 0 | 25.1 | 24.4 | 3.22 | 3.616 |
| 11 | 22.3 | 23 | 10 | 22.8 | 21.9 | 2.85 | 3.609 |
| 12 | 17.4 | 18.1 | 29.8 | 17.7 | 17 | 2.41 | 3.601 |
| 13 | 12.7 | 13 | 49.5 | 12.6 | 12.2 | 1.94 | 3.593 |
| 14 | 24.9 | 25.5 | 25.1 | 0 | 24.5 | 2.37 | 3.575 |
| 15 | 22.4 | 23 | 22.6 | 10.2 | 21.8 | 2.55 | 3.594 |
| 16 | 17.4 | 17.9 | 17.4 | 30.3 | 17 | 2.46 | 3.616 |
| 17 | 12.6 | 12.3 | 12.7 | 50.2 | 12.2 | 2.31 | 3.644 |
| 18 | 10.4 | 10.5 | 10.4 | 9.6 | 59.1 | 2.94 | 3.579 |
| 19 | 2.1 | 2.1 | 2.2 | 2.4 | 91.2 | 1.99 | 3.539 |

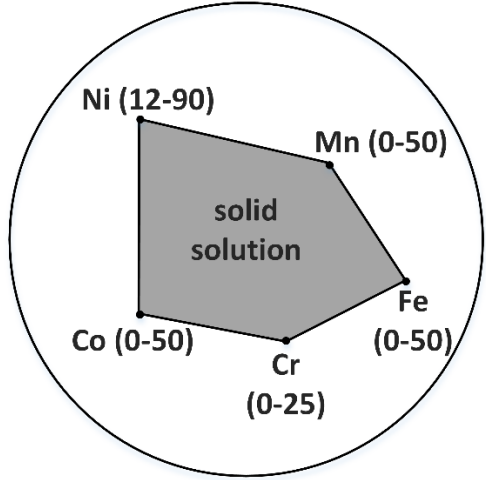

Figure 1. The composition domain which is covered by the experimental dataset in Table 1

## 2. 2. Proposed approach

For explaining the approach, solid-solution binary systems can be considered first. For example, Cu-Ni system with a complete solid solubility can be considered. Figure 2 shows the yield stress of Cu-Ni solid solution alloys versus composition [13]. It can be seen that the yield stress ($\sigma_y$) of solid solution Cu-Ni alloys can be modeled by a polynomial where the polynomial equation can be written as following

$$\sigma_y = 4.449 \times Ni - 0.0321 \times Ni^2 + 0.0064 \times Cu^2 \qquad \text{(Eq. 1)}$$

where *Ni* and *Cu* show the atomic concentration (at.%) of nickel and copper respectively.

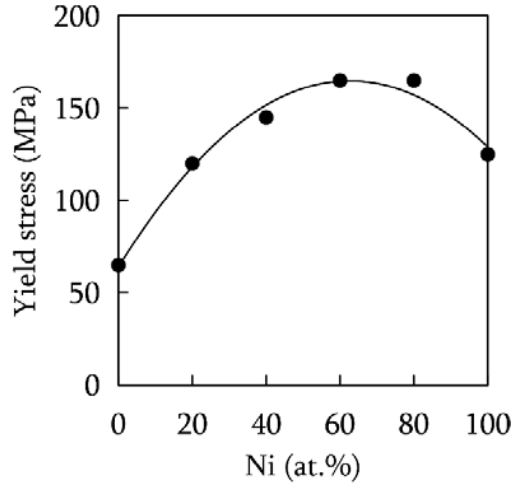

Figure 2. The yield stress (●) of Cu-Ni solid solution alloys versus composition [13] and the proposed polynomial for modeling the strength

In a similar manner, it can be shown that the strength of binary systems with complete solid solubility such as Nb-Ta [14], Mo-Ta [15] or Mo-W [15] can be modeled by polynomials. Furthermore, it may be assumed that the strength of solid-solution HEAs can be modeled by polynomials as well. Therefore, it is proposed in the present work that the hardness ($H$) of solid solution HEAs in Co-Cr-Fe-Mn-Ni system can be modeled by following polynomial

$$H = \sum_{i=1}^{5}(a_i C_i + b_i C_i^2) \tag{Eq. 2}$$

where $a_i$ and $b_i$ are coefficients to be determined and the $C_i$ is the atomic fraction of element *i*. For determining the coefficients, the above equation can be fitted to the experimental dataset in Table 1 by using the least-square method. The obtained values for coefficients are shown in Table 2.

Table 2. The values of coefficients for proposed polynomial

| element | Co | Cr | Fe | Mn | Ni |
|---|---|---|---|---|---|
| $a_i$ | 3.5928 | 7.8797 | -2.2453 | 2.4492 | 5.5452 |
| $b_i$ | -1.9858 | -18.1814 | 4.1906 | -1.8303 | -3.9653 |

## 3. Results and Discussion

The comparison between experimental (Table 1) and predicted results is shown in Figure 3. A good agreement can be seen between predictions and experimental results suggesting that the proposed equation could model the hardness of quaternary and quinary fcc solid solution alloys in Co-Cr-Fe-Mn-Ni system.

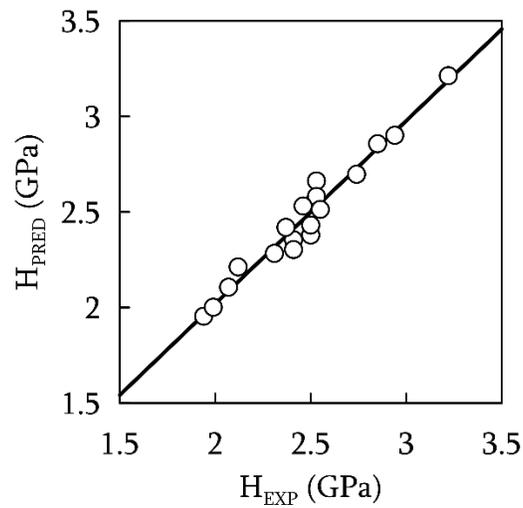

Figure 3. Comparison between the experimental results and predictions

To further evaluate the extrapolating ability of developed polynomial, it is used for predicting the strength of some non-equiatomic alloys in Co-Cr-Fe-Mn-Ni system. The investigated alloys are $Co_xFe_{50-x}Cr_{25}Ni_{25}$ (x = 20, 25, 30, 35) [16], $Cr_xMn_xFe_xCo_xNi_{100-4x}$ (x = 2, 3, 4, 7.5, 10, 12.5, 15, 17.5, 18.75, 20) [12], $Fe_{64-x}Mn_xNi_{28}Co_{5.6}Cr_{2.4}$ (x = 21, 24, 27, 34, 38) [17], $Fe_x(CoCrMnNi)_{100-x}$ (x = 20, 40, 50, 60) [18], $Fe_x(NiCrCo)_{100-x}$ (x = 25, 45, 55) [19], and $CoCrFeMn_xNi_{2-x}$ (x=0.25-1.25) [20]. The comparison between prediction and experimental results are shown in Figure 4. It can be seen that the developed polynomial can predict the changes of strength versus composition relatively well. Surprisingly the prediction results for $Cr_xMn_xFe_xCo_xNi_{100-4x}$ indicate a maximum in hardness with the variation of x in a good agreement with experimental results [12]. For alloy systems $Fe_{64-x}Mn_xNi_{28}Co_{5.6}Cr_{2.4}$ [17] and $CoCrFeMn_xNi_{2-x}$ [20] small changes of hardness versus composition are predicted which is in accordance to the plateau observed for the strength of these alloys [17 and 20]. One point which need to be emphasized is that the relation between the yield stress and nanohardness (GPa) is not necessarily linear [21-22]. In other words, the correlation between nanohardness and yield stress may not be linear [21-22]. This might be the reason for discrepancies observed in Figure 4. According to the results in Figure 4, it can be concluded that the developed polynomial is reliable for predicting the strength of quaternary and quinary solid solution alloys in Co-Cr-Fe-Mn-Ni system.

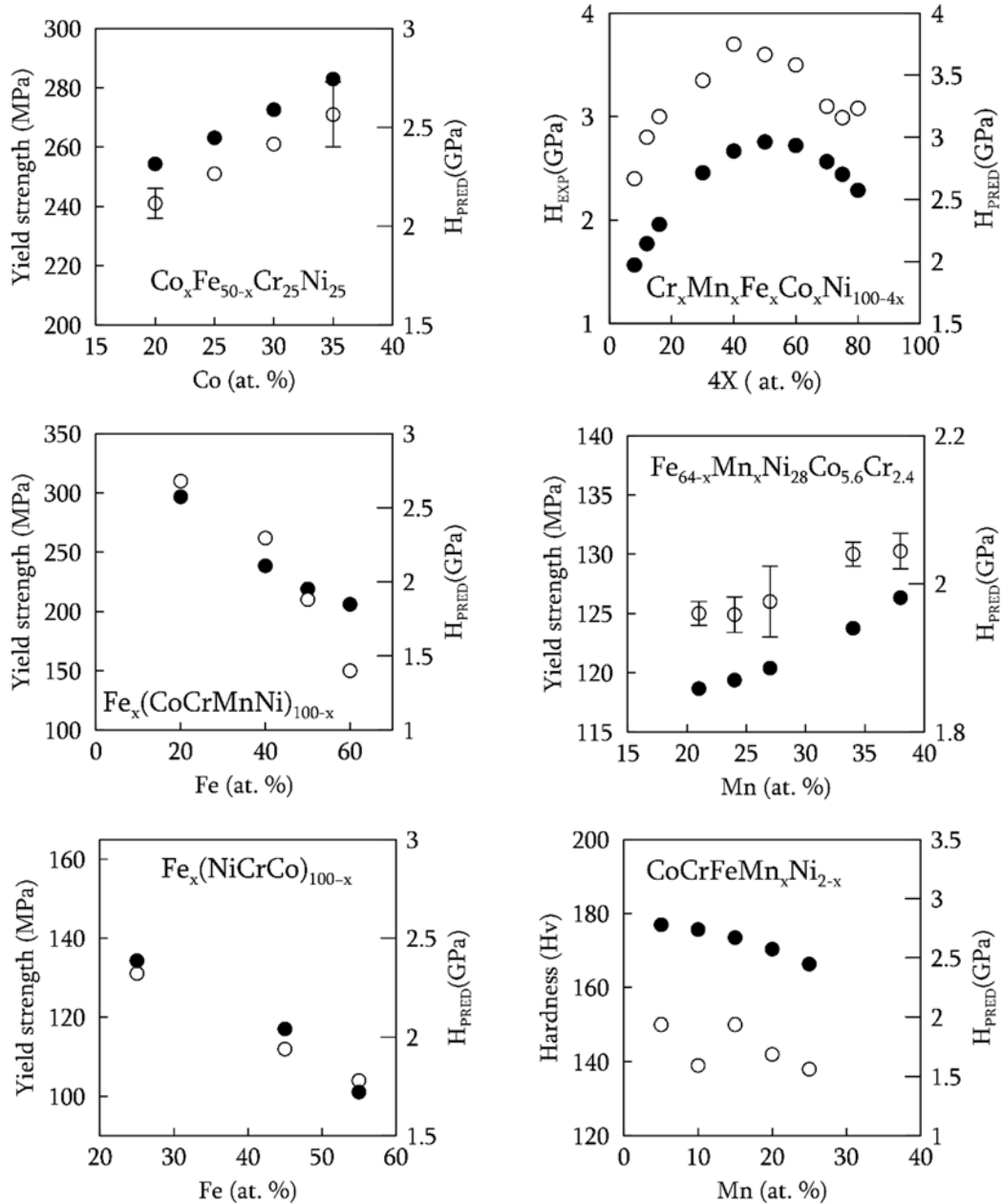

Figure 4. Comparisons between predicted results for hardness (●) and experimental results for strength (○) for a series of non-equiatomic fcc HEAs in Co-Cr-Fe-Mn-Ni system

Because the experimental dataset only includes quaternary and quinary alloys, the polynomial is probably not accurate for binary and ternary sub systems. In other words, when the concentration of two or three elements simultaneously reaches zero, some degree of error should be expected in the prediction results. To investigate this point, experimental data in Table 3 can be considered. As it can be seen, predictions show that the hardness of NiCo is higher than the hardness of alloys NiFe, NiFe-20Cr and NiCoFe which is not in accordance to the experimental results. So the polynomial is definitely not valid for predicting the strength of binary systems. If experimental data can be provided for binary and ternary systems, then the polynomial can be improved for predicting the strength of binary and ternary systems.

Table 3. Single crystal nanoindentation hardness for some of the binary and ternary solid solution alloys in Co-Cr-Fe-Mn-Ni system [23] and prediction results

| Alloy | H (GPa) [23] (experiment) | H(GPa) (prediction) |
| --- | --- | --- |
| NiFe-20Cr | 1.91 | 2.20 |
| NiCoFe | 1.82 | 2.11 |
| NiCoCr | 2.74 | (3.01)* |
| NiCo | 1.31 | 3.08 |
| NiFe | 1.96 | 1.70 |
| Ni | 1.18 | 1.57 |

* The polynomial could not be used for alloy CoCrNi because the Cr concentration of this alloy is not within the Cr concentration of dataset in Table 1

The developed polynomial can be used for investigating the strength of quaternary and quinary alloys in Co-Cr-Fe-Mn-Ni system. Figure 5 shows the predicted results for quaternary alloys $(CoCrMn)_xNi_{100-x}$, $(CoCrFe)_xNi_{100-x}$, $(CoFeMn)_xNi_{100-x}$ and $(CrFeMn)_xNi_{100-x}$. As it can be seen, the hardest alloy in each alloy system have non equiatomic element ratios similar to the experimental results for alloy system $(CoCrFeMn)_xNi_{100-x}$ [12]. Furthermore, it can be seen that the hardness of equiatomic alloys decreases in the following order: CoCrMnNi (3.24 GPa) > CoCrFeNi (2.44 GPa) > CrFeMnNi (2.17 GPa) > CoMnFeNi (2.11 GPa) which is in agreement to the experimental results [24] for the yield stresses: CoCrMnNi ($\sigma_y$ = 282 MPa) > CoCrFeNi ($\sigma_y$ = 274 MPa) > CoMnFeNi ($\sigma_y$ = 176 MPa). These results again confirm the ability of developed polynomial for predicting the strength of HEAs.

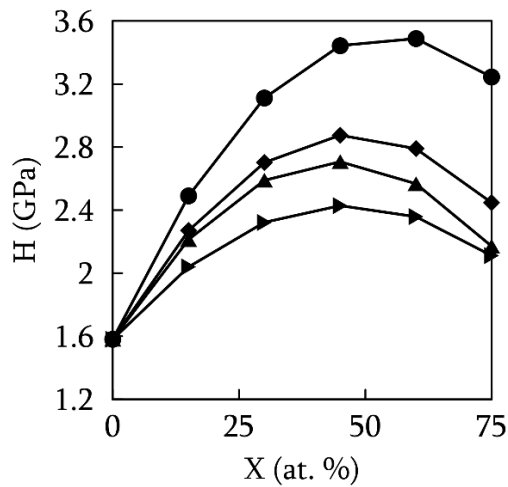

Figure 5. The prediction results for the hardness of alloys $(CoCrMn)_xNi_{100-x}$ (●), $(CoCrFe)_xNi_{100-x}$ (♦), $(CoFeMn)_xNi_{100-x}$ (►) and $(CrFeMn)_xNi_{100-x}$ (▲)

The developed polynomial can be used for investigating the effect of constituent elements on the strength of solid solution alloys. In this regard a MATLAB code is written (Appendix A) and more than 14000 quinary alloys are designed in Co-Cr-Fe-Mn-Ni system. The concentration of constituent elements in designed alloys are selected in the following range: 10 < Co < 50, 10 < Cr < 25, 10 < Fe < 50, 10 < Mn < 50, 15 < Ni < 60. The minimum concentration of Co, Cr, Fe and Mn is considered to be 10 at.% according to the chemical range proposed for HEAs in [12]. Furthermore, the minimum concentration of Ni in designed alloys (15 at.%) is selected to be within the composition domain covered by the experimental dataset (Figure 1).

Figure 6 shows the effect of constituent elements on the hardness of designed quinary alloys. According to the prediction results, it can be seen that by increasing the Fe content, the hardness decreases. The results in Figure 6 suggests that for designing harder fcc solid solution alloys in Co-Cr-Fe-Mn-Ni system, the concentration of Fe should be reduced. In fact, experimental results for alloys $Co_xFe_{50-x}Cr_{25}Ni_{25}$ [16], $Fe_x(CoCrMnNi)_{100-x}$ [18] and $Fe_x(NiCrCo)_{100-x}$ [19] indicate that strength decreases with increasing the Fe content. Furthermore, experimental results [24] indicate that Fe-free equiatomic alloy CoCrMnNi has a higher yield stress in comparison with Fe containing equiatomic alloys CoCrFeMnNi, CoFeMnNi, and CoCrFeNi [24].

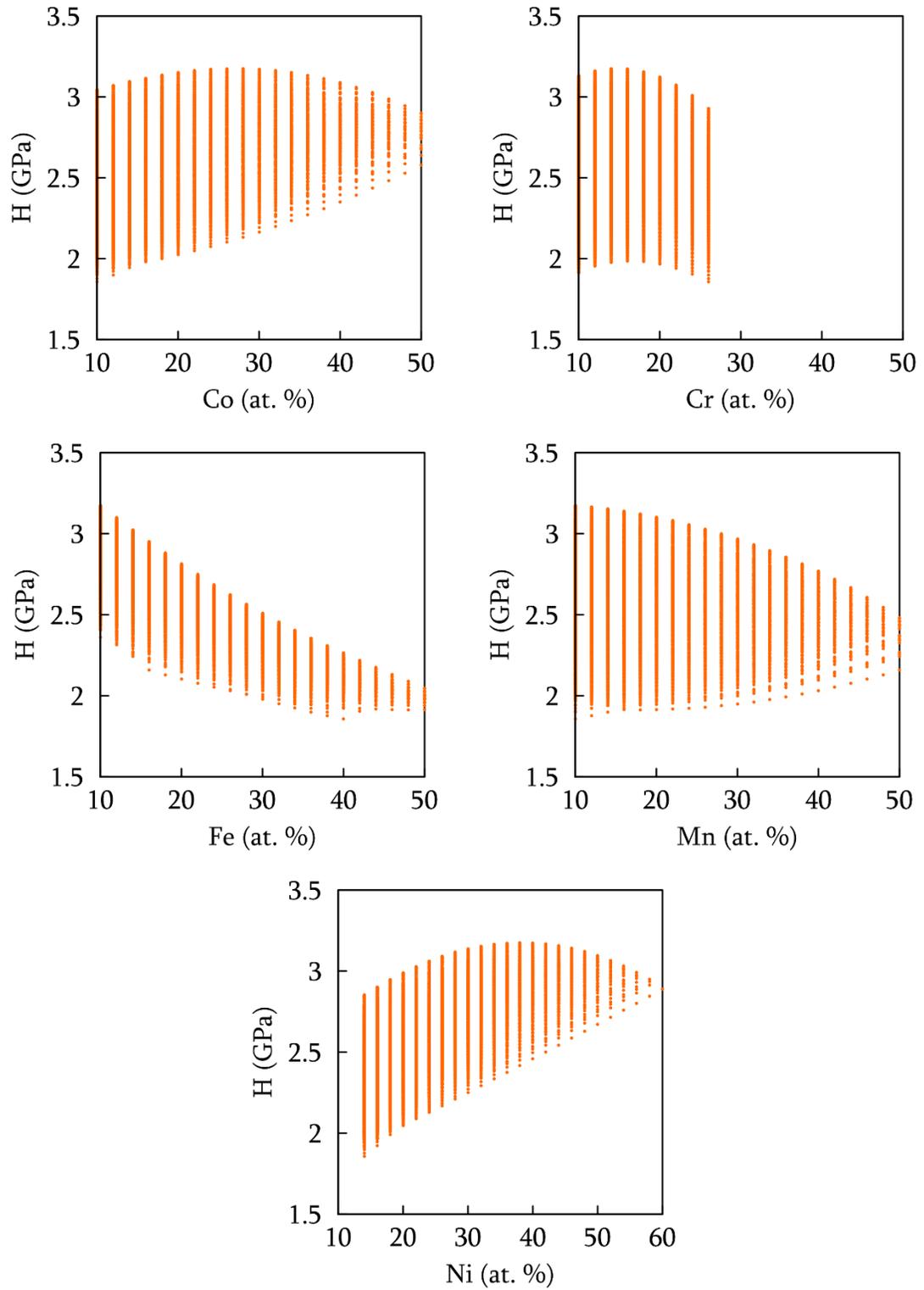

Figure 6. Effects of constituent elements concentration on the hardness ($H$) of designed quinary HEAs in Co-Cr-Fe-Mn-Ni system

It should be noted that the trends which are observed in Figure 6 are only valid for the quinary alloy system Co-Cr-Fe-Mn-Ni and within the investigated composition domain. With changing the alloy system or the alloy concentration the effect of Fe can change. For example, it is well known that when Fe is added to Ni the strength will increase. Therefore, the strengthening or softening effect of Fe depends on the alloy system and the composition range being investigated.

Some of the important parameters which can affect the strength of a solid solution alloy are Peierls-Nabarro (PN) stress [25-28], valence electron concentration [29], atomic size differences [30], lattice distortion [21], electronegativity differences [31], elastic misfits [10], short range orders [32], magnetism [33] and stacking fault energies [34]. Therefore, in order to investigate the effect of an alloying element, all of the above mentioned parameters should be considered. So, the trends which are observed for the constituent elements may be explained by considering the role of a constituent element in changing the above parameters and how changing the above parameters can affect the strength (strengthening or softening).

The effect of Fe on the hardness may be explained by considering its effect on the Peierls-Nabarro (PN) stress or the intrinsic lattice resistance to dislocation motions [18, 25-28]. Several research works have suggested that high yield stresses of solid solution HEAs is due to the enhanced PN stress of these alloys [25-28]. The PN stress for an edge dislocation can be written as

$$\sigma_{PN} = \frac{2G}{1-\nu} \exp(-2\pi\alpha) \qquad \text{(Eq. 3)}$$

where $G$ is the shear modules, $\nu$ is the Poisson ratio, and $\alpha = \xi/b$ where $\xi$ is the dislocation core width and $b$ is the Burgers vector. The parameter $\alpha$ defines the core structure of a dislocation. According to above equation, the PN stress depends both on the interatomic bonds between atoms ($G$ and $v$) and the lattice ($\alpha$). Furthermore, it can be seen that the PN stress (and as a results *H*) decreases with decreasing the shear modules ($G$). It was reported that Fe containing solid solution alloys have in general lower elastic constants [12 and 30]. Furthermore, following elastic constants are proposed for constituent elements in Co-Cr-Fe-Mn-Ni system: $G_{Co}$=81 GPa, $G_{Cr}$=103.5 GPa, $G_{Fe}$= 51.7 GPa, $G_{Mn}$=81 GPa, $G_{Ni}$= 76 GPa [12 and 30]. It can be seen that Fe has the lowest elastic modulus among the constituent elements. Therefore, it can be predicted that the shear modules and, as a result, the PN stress will decrease by the addition of Fe. This may be the reason for decreasing the strength with increasing the Fe content of quinary alloys.

The hardest quinary alloy in Co-Cr-Fe-Mn-Ni system within the investigated composition range is predicted to be $Co_{28}Cr_{14}Fe_{10}Mn_{10}Ni_{38}$ with the hardness of 3.17 GPa. Furthermore, the softest quinary alloy within the investigated composition range is predicted to be $Co_{10}Cr_{26}Fe_{40}Mn_{10}Ni_{14}$ with the hardness of 1.85 GPa. This example shows how the developed polynomial could be used for identifying the softest or hardest alloy in Co-Cr-Fe-Mn-Ni system. A similar approach also could be used for quaternary systems. Therefore, the developed polynomial can be considered as a convenient tool for designing HEAs.

It was proposed that parameters such as lattice distortion, atomic size difference, valence electron concentration, elastic moduli, dislocation widths, electronegativity difference could affect the strength of solid solution HEAs [1-4]. The developed model can be used to assess the relation between these parameters and the strength of solid solution HEAs in Co-Cr-Fe-Mn-Ni. The relation between parameters enthalpy of mixing ($\Delta H_{mix}$) and valence electron concentration (*VEC*) and the strength of alloys is investigated here. These parameters are selected here because their effects on the strength of solid solution alloys is not well investigated. The effect of enthalpy of mixing ($\Delta H_{mix}$) on the microstructure of HEAs is well investigated [3-4], but the effect of $\Delta H_{mix}$ on the strength of HEAs is not inspected so far. So, for the first time in the present work the effect of $\Delta H_{mix}$ on the strength of HEAs is investigated. Mixing enthalpy ($\Delta H_{mix}$) can be regarded as the result of interactions between the constituent elements [35], so it is reasonable to assume that $\Delta H_{mix}$ can affect the strength of atomic bonds and the mechanical behavior of an alloy. The enthalpy of mixing ($\Delta H_{mix}$) for alloys can be estimated by following equation [3-4]:

$$\Delta H_{mix} = \sum_{i=1, i \neq j}^{n}(4\Delta H_{AB}^{mix})C_i C_j \qquad \text{(Eq. 4)}$$

where $\Delta H_{AB}^{mix}$ is the mixing enthalpy of binary liquid alloys and the values can be obtained from [36], $C_i$ is the atomic fraction of element *i* and $C_j$ is the atomic fraction of element j. The relationship between $H$ and $\Delta H_{mix}$ for designed alloys is shown in Figure 7.

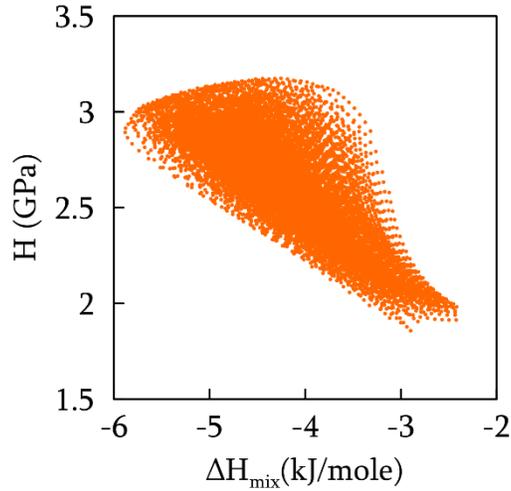

Figure 7. The relation between the enthalpy of mixing ($\Delta H_{mix}$) and the hardness ($H$) of designed quinary HEAs in Co-Cr-Fe-Mn-Ni system

According to this figure, a correlation can be seen between $\Delta H_{mix}$ and $H$, and $H$ increases with decreasing the value of $\Delta H_{mix}$. So it can be concluded that $\Delta H_{mix}$ has important effects on the strength of HEAs and can be considered as a guideline for designing stronger HEAs. In other words, for designing stronger solid solution fcc HEAs, $\Delta H_{mix}$ should be more negative. For example, one may name fcc single solid solution equiatomic CoNiV alloy [37]. The room temperature yield stress of this alloy is reported to be 688 MPa (grain size 8 micron) [37] which is much higher than the yield stress of fcc solid solution alloys in Co-Cr-Fe-Mn-Ni system. On the other hand, $\Delta H_{mix}$ for CoNiV alloy is -14.2 kJ/mol which is much more negative than the $\Delta H_{mix}$ for fcc solid solution alloys in Co-Cr-Fe-Mn-Ni system. So outstanding yield strength of CoNiV alloy may be attributed to its more negative $\Delta H_{mix}$ according to the results in Figure 7. As an another example, one may name fcc single solid solution alloy $Ni_{63.2}V_{36.8}$ [31] with excellent room temperature yield stress of 750 MPa (grain size 8 micron) which is again much higher than the yield stress of fcc solid solution alloys in Co-Cr-Fe-Mn-Ni system. The $\Delta H_{mix}$ for this alloy is -16.745 kJ/mol which is much more negative than the $\Delta H_{mix}$ for alloys in Co-Cr-Fe-Mn-Ni system.

According to the classical model by Miedema [38], more negative values of enthalpy of mixing is due to the increased hybridization because of the larger charge sharing between the atoms [38]. In other words, more negative values of enthalpy of mixing indicate stronger atomic bonds between the constituent elements (and probably higher elastic modules) and this may explain the observed relationships in Figure 7. Furthermore, very recently Oh et al [31] have shown that the governing parameter for solid-solution strengthening in single phase fcc HEAs consisting of 3d transition metal elements is the variation in the atomic-level pressures originating from the charge transfer between neighboring elements [31]. So, considering the effect of charge transfer on the enthalpy of mixing, the relationships observed in Figure 7 may also be explained [31, 38]. In other words, more amount of charge transfer causes more atomic-level pressures leading to the stronger alloys according to [31]. On the other hand, more amount of charge transfer leads to more negative values of enthalpy of mixing according to Miedema's model, and this might be the reason for the relationships observed in Figure 7.

The observed relation between hardness and enthalpy of mixing can also be explained by considering the presence of chemical short-range order (SRO) in solid solution high entropy alloys. It has been shown that the distribution of elements in solid solution HEAs is not completely random, and chemical short-range order (SRO) exists in the structure of these alloys [39-43]. Furthermore, direct atomistic simulations predict that SRO has a negative contribution to the heat of mixing in the order of meV/atom [40-43]. Therefore, with increasing the extent of SRO, enthalpy of mixing decreases and more heat is released during the formation of solution. On the other hand, the presence of SRO can increase the lattice resistance to dislocation motions leading to the hardening of the alloy [44-45]. So, it can be speculated that solid solutions with lower values of $\Delta H_{mix}$ should have a higher degree of SRO, and this could be the reason for increasing the strength with decreasing enthalpy of mixing (Figure 7).

The relation between valance electron concentration (VEC) and the hardness of designed alloys is investigated and the results are shown in Figure 8. Following equation is used for calculating VEC

$$VEC = \sum_{i=1}^{5} C_i \times VEC_i \qquad \text{(Eq. 5)}$$

Where $C_i$ is the atomic fraction of element *i* and $VEC_i$ is valence electron concentration of element *i* values can be found in [3]. According to Figure 8, for designed alloys within the investigated composition hardness increases with increasing $VEC$. In general, high VEC lead to greater atomic bond forces [29], and this may be the reason for the relation observed in Figure 8. Another speculation for the relation observed in Figure 8 could be increasing the directionality of the electron bond forces with increasing valence electron concentration which can increase the PN stress of an alloy [46-47]. Further studies are needed to explain the relationship between hardness and $VEC$.

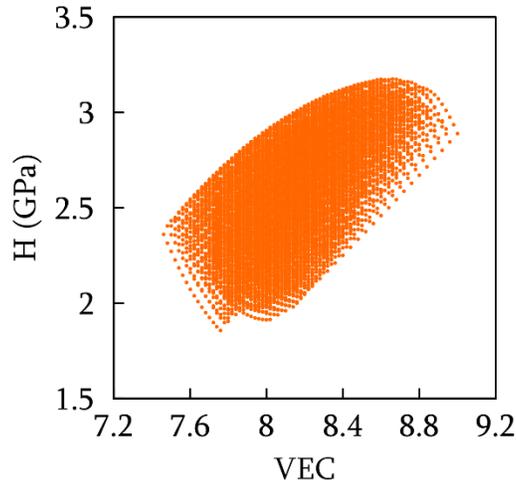

Figure 8. The relation between the valence electron concentration ($VEC$) and the hardness ($H$) of designed quinary HEAs in Co-Cr-Fe-Mn-Ni system

At the end, it should be noted that the developed approach may also be used for modeling other properties of solid solution alloys as well.  For example, the proposed polynomial is applied for modelling the lattice parameter of alloys in Table 1, and the coefficients in Table 4 are obtained after fitting the polynomial to the experimental data. The comparisons between the predictions and experimental data are shown in Figure 9 and an excellent agreement can be seen. Therefore, the approach which is proposed here may also be used for modelling other properties of solid solution alloys as well.

Table 4. The values of coefficients for proposed polynomial to model the lattice parameter of alloys

| element | Co | Cr | Fe | Mn | Ni |
| --- | --- | --- | --- | --- | --- |
| $a_i$ | 3.5346 | 3.5683 | 3.5407 | 3.7094 | 3.6359 |
| $b_i$ | -0.0017 | 0.1541 | 0.0653 | 0.0092 | -0.1118 |

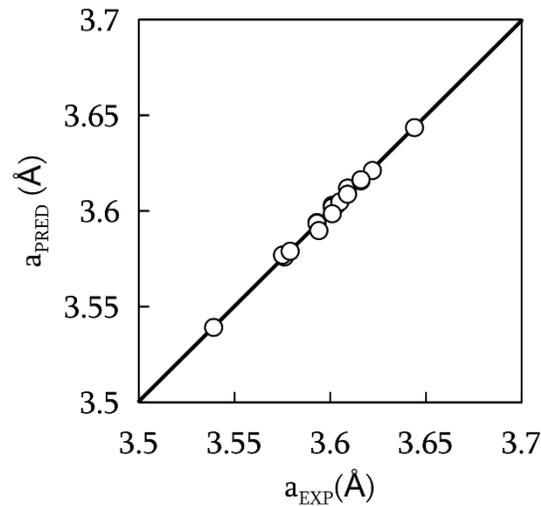

Figure 9. Comparison between the experimental results and predictions for lattice parameter

**Conclusions**

In summary, a simple methodology is presented for predicting the strength of quaternary and quinary fcc solid solution high entropy alloys in the Co-Cr-Fe-Mn-Ni system. It is proposed that the strength of alloys could be modeled by a polynomial of degree 2 where experimental data are used for finding the polynomial's coefficients. The results show that the proposed polynomial could model the hardness of solid solution alloys with a very good accuracy. The proposed polynomial only requires the element concentrations as inputs which makes it very convenient to be applied. The developed polynomial is used for predicting the strength of quinary HEAs in Co-Cr-Fe-Mn-Ni system. The results indicate that with increasing the Fe concentration in designed quinary alloys, the strength of alloys decreases in agreement with experiential data. The softening effect of Fe is explained by considering its effect on decreasing the shear modulus and Peierls-Nabarro (PN) stress of alloys. The effects of enthalpy of mixing and valence electron concentration on the strength of HEAs are investigated. The results show that the enthalpy of mixing has a noticeable effect on the hardness of quinary Co-Cr-Fe-Mn-Ni alloys and the strength of alloys increases with decreasing the enthalpy of mixing. Furthermore, the results show that hardness of quinary Co-Cr-Fe-Mn-Ni alloys increases with increasing the parameter valence electron concentration. Explanations are made to describe the observed relationships.

**References**

[1] E. P. George, D. Raabe, R. O. Ritchie, High-entropy alloys, Nature Reviews Materials, 2019, 4, 515-534

[2] D. B. Miracle, High entropy alloys as a bold step forward in alloy development, Nature Communications, 2019, 10, article number: 1805

[3] D. B. Miracle, O. N. Senkov, A critical review of high entropy alloys and related concepts, Acta Materialia, 2017, 122, 448-511

[4] Y.F. Ye, Q. Wang, J. Lu, C.T. Liu, Y. Yang, High-entropy alloy: challenges and prospects, Materials Today, 2016, 19(6), 349-362

[5] C. Zhang, F. Zhang, S. Chen, W. Cao, Computational thermodynamics aided high-entropy alloy design, JOM, 2012, 64, 839-845

[6] C. Jiang, B.P. Uberuaga, Efficient Ab initio modeling of random multicomponent alloys, Physical Review Letters, 2016, 116, 105501

[7] Y. Lederer, C. Toher, K.S. Vecchio, S. Curtarolo, The search for high entropy alloys: a high throughput ab-initio approach, Acta Materilia, 2018, 159, 364-383

[8] O.N. Senkov, J.D. Miller, D.B. Miracle, C. Woodward, Accelerated exploration of multiprincipal element alloys with solid solution phases, Nature Communications, 2015, 6, 6529

[9] C. Varvenne, A. Luque, W. A. Curtin, Theory of strengthening in fcc high entropy alloys, Acta Materialia, 2016, 118, 164-176

[10] F. Thiel, D. Utt, A. Kauffmann, K. Nielsch, K. Albe, M. Heilmaier, J. Freudenberger, Breakdown of Varvenne scaling in $(AuNiPdPt)_{1-x}Cu_x$ high-entropy alloys, Scripta Materialia, 2020, 18, 15-18

[11] G. Bracq, M. Laurent-Brocq, C. Varvenne, L. Perrière, W.A. Curtin, J. M. Joubert, I. Guillot, Combining experiments and modelling to explore the solid solution strengthening of high and medium entropy alloys, Acta Materialia, 2019, 177, 1, 266-279

[12] M. L. Brocq, L. Perrière, R. Pirès, F. Prima, P. Vermaut, Y. Champion, From diluted solid solutions to high entropy alloys: on the evolution of properties with composition of multi-components alloys, Materials Science and Engineering: A, 2017, 696, 228-235

[13] D.R. Askeland, P.P. Fulay, and W.J. Wright: The Science and Engineering of Materials, 3th ed., Cengage Learning, Stamford, CT, 2011, page 266

[14] B. C. Peters, A. Hendrickson, The strength of tantalum columbium alloy single crystals, Acta Metallurgica, 1966, 14 (9), 1121-1122

[15] J. R. Stephens, W. R. Witzke, Alloy hardening and softening in binary molybdenum alloys as related to electron concentration, Journal of the less common metals, 1972, 29, 371-388

[16] W. Fang, R. Chang, X. Zhang, P. Ji, X. Wang, B. Liu, J. Li, X. He, X. Qu, F. Yin, Effects of cobalt on the structure and mechanical behavior of non-equal molar

$Co_x Fe_{50-x} Cr_{25} Ni_{25}$ high entropy alloys, Materials Science and Engineering: A, 2018, 723, 221-228

[17] K.G. Pradeep, C.C. Tasan, M.J. Yao, Y. Deng, H. Springer, D. Raabe, Non-equiatomic high entropy alloys: approach towards rapid alloy screening and property-oriented design, Materials Science and Engineering: A, 2015, 648, 183-192

[18] M. P. Agustianingrum, I. Ondicho, D. E. Jodi, N. Park, U. Lee, Theoretical evaluation of solid solution interaction in $Fe_x(CoCrMnNi)_{100-x}$ medium- and high-entropy alloys, Materials Science & Engineering A, 2019, 759, 633-639

[19] Y. Zhou, X. Jin, X. Y. Du, L. Zhang, B. S. Li, Comparison of the structure and properties of equiatomic and non-equiatomic multicomponent alloys, Materials Science and Technology, 2018, 34 (8), 988-991

[20] M. Tian, C. Wu, Y. Liu, H. Peng, J. Wang, X. Su, Phase stability and microhardness of $CoCrFeMn_xNi_{2-x}$ high entropy alloys, Journal of Alloys and Compounds, 2019, 811, 152025

[21] M. Laurent-Brocq, L. Perrière, R. Pirès, G. Bracq, T. Rieger, Y. Danard, I. Guillot, Combining tensile tests and nanoindentation to explore the strengthening of high and medium entropy alloys, Materialia, 2019, 7, 100404

[22] F. Gil Coury, P. Wilson, K. D. Clarke, M. J. Kaufman, A. J. Clarke, High-throughput solid solution strengthening characterization in high entropy alloys, Acta Materialia, 2019, 167, 1-11

[23] K. Jin, Y. F. Gao, H. Bei, Intrinsic properties and strengthening mechanism of monocrystalline Ni containing ternary concentrated solid solutions, Materials Science & Engineering A, 2017, 695, 74-79

[24] Z. Wu, H. Bei, G.M. Pharr, E.P. George, Temperature dependence of the mechanical properties of equiatomic solid solution alloys with face-centered cubic crystal structures, Acta Materialia, 2014, 81, 428-441

[25] Y.Y. Zhao, T.G. Nieh, Correlation between lattice distortion and friction stress in Ni-based equiatomic alloys, Intermetallics, 2017, 86, 45-50

[26] X. Liu, Z. Pei, M. Eisenbach, Dislocation core structures and Peierls stresses of the high-entropy alloy NiCoFeCrMn and its subsystems, Materials and Design, 2019, 180, 107955

[27] S. S. Sohn, A. K. Silva, Y. Ikeda, F. Körmann, W. Lu, W. S. Choi, B. Gault, D. Ponge, J. Neugebauer, D. Raabe, Ultrastrong medium-entropy single-phase alloys designed via severe lattice distortion, Advanced Materials, 2019, 31 (8), 1807142

[28] L. Zhang, Y. Xiang, J. Han, D. J. Srolovitz, The effect of randomness on the strength of high-entropy alloys, Acta Materialia, 2019, 166, 424-434

[29] R. Chen, G. Qin, H. Zheng, L. Wang, Y. Su, Y. L. Chiu, H. Ding, J. Guo, H. Fu, Composition design of high entropy alloys using the valence electron concentration to balance strength and ductility, Acta Materialia, 2018, 144, 129-137

[30] F. G. Coury, K. D. Clarke, C. S. Kiminami, M. J. Kaufman, A. J. Clarke, High throughput discovery and design of strong multicomponent metallic solid solutions, Scientific Reports, 2018, 8, 8600

[31] H. S. Oh, S. J. Kim, K. Odbadrakh, W. H. Ryu, K. N. Yoon, S. Mu, F. Körmann, Y. Ikeda, C. C. Tasan, D. Raabe, T. Egami, E. S. Park, Engineering atomic-level complexity in high-entropy and complex concentrated alloys, Nature Communications, 2019, 10, 2090

[32] Q. J. Li, H. Sheng, E. Ma, Strengthening in multi-principal element alloys with local-chemical-order roughened dislocation pathways, Nature Communications, 2019, 10, 3563

[33] C. Niu, C. R. LaRosa, J. Miao, M. J. Mills, M. Ghazisaeidi, Magnetically-driven phase transformation strengthening in high entropy alloys, Nature Communications, 2018, 9, 1363

[34] Y. Zeng, X. Cai, M. Koslowski, Effects of the stacking fault energy fluctuations on the strengthening of alloys, Acta Materialia, 2019, 164, Pages 1-11

[35] M.C. Gao, J. W. Yeh, P.K. Liaw, Y. Zhang, High-entropy alloys: fundamentals and applications, Springer Publishing Co., New York, NY (2016)

[36] A.Takeuchi and A.Inoue, Classification of bulk metallic glasses by atomic size difference, heat of mixing and period of constituent elements and its application to characterization of the main alloying element, Materials Transactions, 2005, 46 (12), 2817-2829

[37] S. S. Sohn, A. K. Silva, Y. Ikeda, F. Körmann, W. Lu, W. S. Choi, B. Gault, D. Ponge, J. Neugebauer, D. Raabe, Ultrastrong medium-entropy single-phase alloys designed via severe lattice distortion, Advanced Materials, 2019, 31 (8), 1807142

[38] A.R.Miedema, P.F.de Châtel, F.R.de Boer, Cohesion in alloys - fundamentals of a semi-empirical model, Physica B+C, 1980, 100 (1), 1-28

[39] L. Zhang, Y. Xiang, J. Han, D. J. Srolovitz, The effect of randomness on the strength of high-entropy alloys, Acta Materialia, 2019, 166, 424-434

[40] P. Singh, A. V. Smirnov, D. D. Johnson, Atomic short-range order and incipient long-range order in high entropy alloys, Physical Review B, 2015, 91, 224204

[41] A. Tamm, A. Aabloo, M. Klintenberg, M. Stocks, A. Caro, Atomic-scale properties of Ni-based FCC ternary, and quaternary alloys, Acta Materilia, 2015, 99, 307-312

[42] A. Sharma, P. Singh, D. D. Johnson, P. K. Liaw, G. Balasubramanian, Atomistic clustering-ordering and high-strain deformation of an $Al_{0.1}CrCoFeNi$ high-entropy alloy, Scientific Reports, 2016, 6, 31028

[43] Y. Ma, Q. Wang, C. Li, L. J.Santodonato, M. Feygenson, C. Dong, P. K.Liaw, Chemical short-range orders and the induced structural transition in high-entropy alloys, Scripta Materialia, 2018, 144, 64-68

[44] F. Pettinari-Sturmel, M. Jouiad, H. O. K. Kirchner, N. Clement, A. Coujou, Local disordering and reordering phenomena induced by mobile dislocations in short-range-ordered solid solutions, Philosophical Magazine A, 2002, 82, 3045-3054

[45] F. Pettinari, M. Prem, G. Krexner, P. Caron, A. Coujou, H.O.K Kirchner, N. Clement, Local order in industrial and model γ phases of superalloys, Acta Materialia, 2001, 13(1), 2549-2556

[46] J. R. Stephens, W. R. Witzke, The role of electron concentration in softening and hardening of ternary molybdenum alloys, Journal of the Less Common Metals, 1975, 41 (2), 265-282

[47] Y. Hiraoka, T. Ogusu, N. Yoshizawa, Decrease of yield strength in molybdenum by adding small amounts of Group VIII elements, Journal of Alloys and Compounds, 2004, 381(1-2), 192-196

## Appendix A

The iterative MATLAB code written for designing quinary alloys in Co-Cr-Fe-Mn-Ni system

```
clear all
t=0;
u=0;

vecCo=9;vecCr=6;vecFe=8;vecMn=7;vecNi=10;

HCoCr=-4;HCoFe=-1;HCoMn=-5;HCoNi=0;
HCrFe=-1;HCrMn=2;HCrNi=-7;
HFeMn=0;HFeNi=-2;
HMnNi=-8;

for co=10:2:50
    for cr=10:2:26
        for fe=10:2:50
            for mn=10:2:50
                for ni=14:2:60

                    sum=co+cr+fe+mn+ni;

                    if sum==100

                        u=u+1;
                        comp(u,1)=co; comp(u,2)=cr; comp(u,3)=fe; comp(u,4)=mn; comp(u,5)=ni;

       H1=3.5928*co/100+7.8797*cr/100-
2.2453*fe/100+2.4492*mn/100+5.5452*ni/100;
       H2=-1.9858*co*co/10000-18.1814*cr*cr/10000+4.1906*fe*fe/10000-
1.8303*mn*mn/10000-3.9653*ni*ni/10000;
       H(u,1)=H1+H2;

      Hmix1=4*(HCoCr*co*cr+HCoFe*co*fe+HCoMn*co*mn+HCoNi*co*ni)/10000;
      Hmix2=4*(HCrFe*cr*fe+HCrMn*cr*mn+HCrNi*cr*ni)/10000;
      Hmix3=4*(HFeMn*fe*mn+HFeNi*fe*ni)/10000;
      Hmix4=4*(HMnNi*mn*ni)/10000;
      Hmix(u,1)=Hmix1+Hmix2+Hmix3+Hmix4;

      VEC(u,1)=(co*vecCo+cr*vecCr+fe*vecFe+mn*vecMn+ni*vecNi)/100;

                    end
                end
            end
        end
    end
end

[val, idx] = max(H); [val2, idx2] = min(H);
```